\documentclass[preprintnumbers,amsmath,amssymb,floatfix,12pt,prd,superscriptaddress,nofootinbib]{revtex4}
\usepackage{graphicx}
\usepackage{epsfig}
\usepackage{bm}
\usepackage{amsfonts}

\begin{document}


\title{Interacting entropy-corrected new agegraphic dark energy in Brans-Dicke cosmology}

\author{K. Karami}
\email{KKarami@uok.ac.ir}
 \affiliation{Department of Physics, University
of Kurdistan, Pasdaran St., Sanandaj, Iran } \affiliation{Research
Institute for Astronomy $\&$ Astrophysics of Maragha (RIAAM),
Maragha, Iran}

\author{A. Sheykhi}
\email{sheykhi@mail.uk.ac.ir} \affiliation{Department of Physics,
Shahid Bahonar University, P.O. Box 76175, Kerman,
Iran}\affiliation{Research Institute for Astronomy $\&$ Astrophysics
of Maragha (RIAAM), Maragha, Iran}

\author{M. Jamil}
\email{mjamil@camp.nust.edu.pk} \affiliation{Center for Advanced
Mathematics and Physics (CAMP), National University of Sciences and
Technology (NUST), Islamabad, Pakistan}\affiliation{The Abdus Salam
International Center for Theoretical Physics (ICTP), Strada Costiera
11, Trieste, Italy}

\author{Z. Azarmi}
\email{azarmi_z@yahoo.com}  \affiliation{Department of Physics,
University of Kurdistan, Pasdaran St., Sanandaj, Iran }

\author{M.M. Soltanzadeh}
\email{msoltanzadeh@uok.ac.ir}  \affiliation{Department of Physics,
University of Kurdistan, Pasdaran St., Sanandaj, Iran }

\date{\today}

\begin{abstract}
\centerline{\bf Abstract} Motivated by a recent work of one of us
\cite{Sheykhi2}, we extend it by using quantum (or entropy)
corrected new agegraphic dark energy in the Brans-Dicke cosmology.
The correction terms are motivated from the loop quantum gravity
which is one of the competitive theories of quantum gravity. Taking
the non-flat background spacetime along with the conformal age of
the universe as the length scale, we derive the dynamical equation
of state of dark energy and the deceleration parameter. An important
consequence of this study is the phantom divide scenario with
entropy-corrected new agegraphic dark energy. Moreover, we assume a
system of dark matter, radiation and dark energy, while the later
interacts only with dark matter. We obtain some essential
expressions related with dark energy dynamics. The cosmic
coincidence problem is also resolved in our model.
\end{abstract}



\maketitle

\newpage
\section{Introduction}
The present acceleration of the universe expansion has been well
established through numerous and complementary cosmological
observations \cite{Rie}. A component which is responsible for this
accelerated expansion usually dubbed ``dark energy'' (DE). The
combined analysis of  astronomical observations indicate that
nearly three quarters of the universe consists of DE with negative
pressure. Up to now different DE models have been proposed to
explain this acceleration while most of them cannot explain all
the features of universe or they have so many parameters, making
them difficult to fit. The dynamical nature of DE, at least in an
effective level, can originate from various fields, although a
complete description requires a deeper understanding of the
underlying theory of quantum gravity. Nevertheless, physicists can
still make some attempts to probe the nature of DE according to
some basic quantum gravitational principles. Two examples of such
a paradigm are the holographic DE (HDE) and the agegraphic DE
(ADE) models which are originated and possess some significant
features of quantum gravity. The former, that arose a lot of
enthusiasm recently \cite{Coh,jamil,wang,karami1}, is motivated
from the holographic hypothesis \cite{Suss1} and has been tested
and constrained by various astronomical observations \cite{Xin}.
The later is originated form uncertainty relation of quantum
mechanics together with the gravitational effect in general
relativity. The ADE model assumes that the observed DE comes from
the spacetime and matter field fluctuations in the universe.
Following the line of quantum fluctuations of spacetime,
Karolyhazy et al. \cite{Kar1} discussed that the distance $t$ in
Minkowski spacetime cannot be known to a better accuracy than
$\delta{t}=\lambda t_{P}^{2/3}t^{1/3}$ where $\lambda$ is a
dimensionless constant of order unity and $t_P$ is the reduced
Planck time. Based on Karolyhazy relation, Maziashvili \cite{Maz}
discussed that the energy density of metric fluctuations of the
Minkowski spacetime is given by
\begin{equation}
\rho_{\Lambda}\sim \frac{1}{t_P^2t^2}\sim
\frac{M_P^2}{t^2},\label{rhoMaz}
\end{equation}
where $M_P$ is the reduced Planck mass $M_P^{-2}=8\pi G$. Based on
Karolyhazy relation \cite{Kar1} and Maziashvili arguments
\cite{Maz}, Cai proposed the original ADE model to explain the
accelerated expansion of the universe \cite{Cai1}. The original
ADE has the following energy density \cite{Cai1}
\begin{equation}
\rho_{\Lambda}=\frac{3{n}^2M_P^2}{T^2},\label{OADE}
\end{equation}
where $T$ is the age of the universe and given by
\begin{equation}
T=\int{\rm d}t=\int_0^a\frac{{\rm d}a}{Ha}.
\end{equation}
Also the numerical factor 3$n^2$ is introduced to parameterize
some uncertainties, such as the species of quantum fields in the
universe, the effect of curved spacetime (since the energy density
is derived for Minkowski spacetime), and so on. However, the
original ADE model had some difficulties. For example it suffers
from the difficulty to describe the matter-dominated epoch.
Therefore, a new model of ADE was proposed by Wei and Cai
\cite{Wei2}, while the time scale is chosen to be the conformal
time instead of the age of the universe. The energy density of the
new ADE (NADE) is given by \cite{Wei2}
\begin{equation}
\rho_{\Lambda}=\frac{3{n}^2M_P^2}{\eta^2},\label{NADE}
\end{equation}
where $\eta$ is conformal time of the FRW universe and given by
\begin{equation}
\eta=\int\frac{{\rm d}t}{a}=\int_0^a\frac{{\rm
d}a}{Ha^2}.\label{eta}
\end{equation}
The NADE contains some new features different from the original
ADE and overcome some unsatisfactory points. The ADE models have
been examined and studied in ample detail
\cite{age,shey1,shey2,karami2,karami3}.

It is worthy  to note that the definition and derivation of HDE
density depends on the entropy-area relationship of black holes in
Einstein's gravity \cite{Coh}. However, this definition can be
modified from the inclusion of quantum effects, motivated from the
loop quantum gravity (LQG). The quantum corrections provided to
the entropy-area relationship leads to the curvature correction in
the Einstein-Hilbert action and vice versa \cite{Zhu,Suj}. The
corrected entropy takes the form \cite{Zhang}
\begin{equation}\label{S}
S=\frac{A}{4G}+\tilde{\alpha} \ln {\frac{A}{4G}}+\tilde{\beta},
\end{equation}
where $\tilde{\alpha}$ and $\tilde{\beta}$ are dimensionless
constants of order unity. These corrections arise in the black
hole entropy in LQG due to thermal equilibrium fluctuations and
quantum fluctuations \cite{Rovelli}. Motivated by the corrected
entropy-area relation (\ref{S}), Wei \cite{wei} proposed
 the energy density of the so-called ``entropy-corrected HDE''
 (ECHDE) as
\begin{equation}\label{rhoS}
\rho _{\Lambda}=3n^2M_{P}^{2}L^{-2}+\alpha L^{-4}\ln
(M_{P}^{2}L^{2})+\beta L^{-4},
\end{equation}
where $\alpha$ and $\beta$ are dimensionless constants of order
unity. On the other hand, soon after ``general theory of
relativity" was introduced by Albert Einstein in $1915$, several
attempts to construct alternative theories of gravity were made.
One of the most studied alternative theories was scalar-tensor
theory, where the gravitational action contains, apart from the
metric, a scalar field which describes part of the gravitational
field. The scalar-tensor theory was invented first by Jordan
\cite{Jor} in the 1950's, and then taken over by  Brans and Dicke
\cite{BD} some years later. The starting point of the
scalar-tensor theory sharing stage with gravitation is the idea of
Mach. This theory has got a new impetus recently as it arises
naturally as the low energy limit of many theories of quantum
gravity such as superstring theory or Kaluza-Klein theory
\cite{Wit1}. Because the ADE density belongs to a dynamical
cosmological constant, we need a dynamical frame to accommodate it
instead of Einstein gravity. Therefore the investigation on the
agegraphic models of DE in the framework of Brans-Dicke theory is
of great importance. The investigation on the
holographic/agegraphic models of DE in the framework of
Brans-Dicke cosmology, have been carried out in
\cite{other,Setare2,Pavon2,Sheykhi1,Xu}.

In this paper we would like to study the so-called
entropy-corrected NADE (ECNADE) whose $L$ in Eq. (\ref{rhoS}) is
replaced with the conformal time $\eta$ of the universe. Therefore
we propose the energy density of ECNADE of the form
\begin{equation}
\rho_{\Lambda} = \frac{3n^2{M_P^2}}{\eta^2} +
\frac{\alpha}{{\eta}^4}\ln{({M_P^2}{\eta}^2)} +
\frac{\beta}{\eta^4}.\label{ecnade}
\end{equation}
In the special case $\alpha=\beta=0$, Eq. (\ref{ecnade}) yields
the NADE density (\ref{NADE}) in Einstein gravity \cite{Wei2}.
Note that in Eq. (\ref{ecnade}) like the NADE density (\ref{NADE})
to justify the matter-dominated era, the time scale is chosen to
be the conformal time instead of the age of the universe.

The motivation idea for taking the energy density of modified NADE
in the form (\ref{ecnade}) comes from the fact that both NADE and
HDE models have the same origin. Indeed, it was argued that the
NADE models are the HDE model with different IR length scales
\cite{Myung}. Since the last two terms in Eq. (\ref{ecnade}) can
be comparable to the first term only when $\eta$ is very small,
the corrections make sense only at the early stage of the
universe. When the time scale $\eta$ becomes large, ECNADE reduces
to the ordinary NADE. Although it is believed that our universe is
flat, a contribution to the Friedmann equation from spatial
curvature is still possible if the number of e-foldings is not
very large (see Huang and Li in \cite{Coh}). Besides, some
experimental data has implied that our universe is not a perfectly
flat universe and recent papers have favored the universe with
spatial curvature \cite{spe}.

\section{ECNADE in Brans-Dicke theory}\label{ECNADE}
The action of four-dimensional Brans-Dicke theory in the canonical
form can be written as \cite{Arik}
\begin{equation}
S=\int {\rm
d}^4x\sqrt{g}\Big(-\frac{1}{8\omega}\phi^2R+\frac{1}{2}g^{\mu\nu}\partial_\mu\phi\partial_\nu\phi+L_M\Big),\label{action}
\end{equation}
where ${R}$ is the scalar curvature and $\phi$ is the Brans-Dicke
scalar field. The non-minimal coupling term $\phi^2 R$ replaces
with the Einstein-Hilbert term ${R}/{G}$ in such a way that
$G^{-1}_{\mathrm{eff}}={2\pi \phi^2}/{\omega}$  where
$G_{\mathrm{eff}}$ is the effective gravitational constant as long
as the dynamical scalar field $\phi$ varies slowly. The signs of
the non-minimal coupling term and the kinetic energy term are
properly adopted to $(+---)$ metric signature. The ECNADE model
will be accommodated in the non-flat Friedmann-Robertson-Walker
(FRW) universe which is described by the line element
\begin{equation}
{\rm d}s^2={\rm d}t^2-a^2(t)\left(\frac{{\rm
d}r^2}{1-kr^2}+r^2{\rm d}\Omega^2\right),\label{metric}
\end{equation}
where $a(t)$ is the scale factor, and $k$ is the curvature
parameter with $k = -1, 0, 1$ corresponding to open, flat, and
closed universes, respectively. A closed universe with a small
positive curvature ($\Omega_k\simeq0.02$) is compatible with
observations \cite{spe}. The field equations can be obtained by
varying action (\ref{action}) with respect to metric
(\ref{metric}) which leads
\begin{equation}
\frac{3}{4\omega}\phi^2\Big(H^2+\frac{k}{a^2}\Big)-\frac{1}{2}\dot{\phi}^2+\frac{3}{2\omega}H\dot{\phi\phi}=\rho_r+\rho_m+\rho_\Lambda,\label{BD1}
\end{equation}
\begin{equation}
\frac{-1}{4\omega}\phi^2\Big(2\frac{\ddot{a}}{a}+H^2+\frac{k}{a^2}\Big)-\frac{1}{\omega}H\dot{\phi\phi}-
\frac{1}{2\omega}\ddot{\phi}\phi-\frac{1}{2}\Big(1+\frac{1}{\omega}\Big)\dot{\phi}^2=p_r+p_\Lambda,\label{BD2}
\end{equation}
\begin{equation}
\ddot{\phi}+3H\dot{\phi}-\frac{3}{2\omega}\Big(\frac{\ddot{a}}{a}+H^2+\frac{k}{a^2}\Big)\phi=0,\label{BD3}
\end{equation}
where the dot stands for the derivative with respect to time and
$H=\dot{a}/a$ is the Hubble parameter. Here $\rho_\Lambda$,
$p_\Lambda$, $\rho_r$, $p_r$ and $\rho_m$ are, respectively, the
DE density, DE pressure, energy density of radiation, pressure of
radiation and energy density of pressureless dust (dark matter).
Here the contribution of radiation is also considered. Because
according to Eq. (\ref{ecnade}), the last two terms make only
sense at the early stage of the universe, i.e. the
radiation-dominated epoch.

At this point our system of equations is not closed and we still
have freedom to choose one. We shall assume that Brans-Dicke field
can be described as a power law of the scale factor, $\phi\propto
a^{\varepsilon}$. In principle there is no compelling reason for
this choice. However, it has been shown that for small
$\varepsilon$ it leads to consistent results \cite{Pavon2,Xu}. A
case of particular interest is that when $\varepsilon$ is small
whereas $\omega$ is high so that the product $\varepsilon \omega$
results of order unity \cite{Pavon2}. This is interesting because
local astronomical experiments set a very high lower bound on
$\omega$; in particular, the Cassini experiment implies that
$\omega>10^4$ \cite{Bert}. Taking the derivative with respect to
time of relation $\phi\propto a^{\varepsilon}$, we get
\begin{equation}
\dot{\phi}=\varepsilon H\phi,\label{phid}
\end{equation}
\begin{equation}
\ddot{\phi}=\varepsilon^2H^2
\phi+\varepsilon\phi\dot{H}.\label{phiddot}
\end{equation}
In the framework of Brans-Dicke cosmology, we write down the
energy density of the ECNADE model in the universe as
\begin{equation}
\rho_\Lambda=\frac{3n^2\phi^2}{4\omega\eta^2}+\frac{\alpha}{\eta^4}\ln\Big(\frac{\phi^2\eta^2}{4\omega}\Big)+\frac{\beta}{\eta^4},\label{ecbd1}
\end{equation}
which can be rewritten as
\begin{equation}
\rho_\Lambda=\frac{3n^2\phi^2}{4\omega\eta^2}\gamma_n,\label{ecbd2}
\end{equation}
where
\begin{equation}
\gamma_n=1+\frac{4\omega\alpha}{3n^2\phi^2\eta^2}\ln
\Big(\frac{\phi^2\eta^2}{4\omega}\Big)+\frac{4\omega\beta}{3n^2\phi^2\eta^2},\label{gamma}
\end{equation}
shows the deviation from the NADE model. In the above equation
$\phi^2={\omega}/{2\pi G_{\mathrm{eff}}}$. In the limiting case
$G_{\mathrm{eff}}\rightarrow G$, expression (\ref{ecbd1}) restores
the energy density of ECNADE in  Einstein gravity \cite{karami3}.
The critical energy density, $\rho_{\mathrm{cr}}$, and the energy
density of the curvature, $\rho_k$, are defined as
\begin{equation}
\rho_{\rm
cr}=\frac{3\phi^2H^2}{4\omega},~~~~~~\rho_k=\frac{3k\phi^2}{4\omega
a^2}.\label{critical}
\end{equation}
The fractional energy densities are also defined as usual
\begin{equation}
\Omega_r=\frac{\rho_r}{\rho_{\rm
cr}}=\frac{4\omega\rho_r}{3\phi^2H^2},~~~\Omega_m=\frac{\rho_m}{\rho_{\rm
cr}}=\frac{4\omega\rho_m}{3\phi^2H^2},~~~\Omega_k=\frac{\rho_k}{\rho_{\rm
cr}}=\frac{k}{H^2a^2},~~~\Omega_\Lambda=\frac{\rho_\Lambda}{\rho_{\rm
cr}}=\frac{n^2}{H^2\eta^2}\gamma_n.\label{omega}
\end{equation}

\subsection{Noninteracting case}
Consider the FRW universe filled with ECNADE, pressureless matter
 and radiation which evolves according to their conservation laws
\begin{equation}
\dot{\rho}_{\Lambda}+3H(1+w_{\Lambda})\rho_{\Lambda}=0,\label{eqde}
\end{equation}
\begin{equation}
\dot{\rho}_{m}+3H\rho_{m}=0,\label{eqm}
\end{equation}
\begin{equation}
\dot{\rho}_{r}+4H\rho_{r}=0,\label{eqr1}
\end{equation}
where $w_{\Lambda}=p_{\Lambda}/\rho_{\Lambda}$ is the equation of
state (EoS) parameter of ECNADE. Taking the derivative of Eq.
(\ref{ecbd2}) with respect to the cosmic time and using Eqs.
(\ref{phid}), (\ref{omega}) and $\dot{\eta}=1/a$ we have
\begin{equation}
\dot{\rho}_\Lambda=2H\rho_\Lambda\left\{\frac{\varepsilon
}{\gamma_n}+\frac{1}{na}\Big(\frac{1}{\gamma_n}-2\Big)\Big(\frac{\Omega_\Lambda}{\gamma_n}\Big)^{1/2}+\frac{4\alpha\omega
H^2}{3n^4\phi^2}\frac{\Omega_\Lambda}{\gamma_n^2}\Big[\varepsilon+\frac{1}{na}\Big(\frac{\Omega_\Lambda}{\gamma_n}\Big)^{1/2}\Big]\right\}.\label{rhod}
\end{equation}
Inserting this equation in the conservation law (\ref{eqde}), we
obtain the EoS parameter of ECNADE model in the framework of
Brans-Dicke theory
\begin{equation}
w_\Lambda=-1-\frac{2\varepsilon}{3}\frac{1}{\gamma_n}+\frac{2}{3na}\Big(2-\frac{1}{\gamma_n}\Big)\Big(\frac{\Omega_\Lambda}{\gamma_n}\Big)^{1/2}
-\frac{8\alpha\omega
H^2}{9n^4\phi^2}\frac{\Omega_\Lambda}{\gamma_n^2}\Big[\varepsilon+\frac{1}{na}
\Big(\frac{\Omega_\Lambda}{\gamma_n}\Big)^{1/2}\Big].\label{eos1}
\end{equation}
In the special case $\alpha=\beta=0$, from Eq. (\ref{gamma}) we
have $\gamma_n=1$ and Eq. (\ref{eos1}) in the absence of radiation
restores the EoS parameter of NADE in Brans-Dicke theory
\cite{Sheykhi2}
\begin{eqnarray}
w_\Lambda=-1-\frac{2\varepsilon}{3}+\frac{2}{3na}\sqrt{\Omega_\Lambda}\label{wDo}.
\end{eqnarray}
Comparing Eq. (\ref{eos1}) with (\ref{wDo}) we see that in the
presence of correction terms the scalar field $\phi$ enters the
EoS parameter explicitly. From Eqs. (\ref{ecbd1}) and (\ref{eos1})
we see that in the late time where $\Omega_\Lambda \rightarrow 1$
and $a \rightarrow\infty$ we can neglect the last two terms in Eq.
(\ref{eos1}) and we find
$w_\Lambda=-1-\frac{2\varepsilon}{3\gamma_n}$. Thus in the late
time universe, although the EoS parameter of ECNADE does not feel
the presence of the last two correction terms in Eq.
(\ref{ecnade}) but for $\varepsilon\neq 0$ it will necessary cross
the phantom divide, i.e. $w_\Lambda<-1$ in Brans-Dicke theory.
This is in contrast to Einstein gravity
($\varepsilon\rightarrow0$) where $w_\Lambda$ of ECNADE mimics a
cosmological constant in the late time \cite{karami3}.

Since in our model the dynamics of the scale factor is governed
not only by the dark matter (DM), the radiation and the NADE, but
also by the Brans-Dicke field, the signature of the deceleration
parameter,
\begin{equation}
q=-\frac{\ddot{a}}{aH^2}=-1-\frac{\dot{H}}{H^2},\label{dec1}
\end{equation}
has to be examined carefully. When deceleration parameter is
combined with the Hubble parameter and the dimensionless density
parameters, form a set of useful parameters for the description of
the astrophysical observations. Dividing  Eq. (\ref{BD2}) by
$H^2$, and using Eqs. (\ref{phid}), (\ref{phiddot}), (\ref{ecbd1})
and (\ref{omega}) we obtain
\begin{equation}
q=\frac{1}{2\varepsilon+2}[(2\varepsilon+1)^2+2\varepsilon(\varepsilon\omega-1)+\Omega_k+3\Omega_\Lambda
w_\Lambda+\Omega_r],\label{dec2}
\end{equation}
where $w_{\Lambda}$ is given by Eq. (\ref{eos1}).

We can also obtain the equation of motion for $\Omega_{\Lambda}$.
Taking the derivative of the last Eq. (\ref{omega}) and using
relation ${\dot{\Omega}_{\Lambda}}=H{\Omega'_{\Lambda}}$, we
obtain
\begin{equation}
\Omega^{'}_{\Lambda}=\Omega_{\Lambda}\left[-\frac{2\dot{H}}{H^2}-\frac{2}{na}\Big(\frac{\Omega_{\Lambda}}{\gamma_n}\Big)^{1/2}+\frac{1}{H}\frac{\dot{\gamma_n}}{\gamma_n}\right],\label{eqmotion}
\end{equation}
where the prime denotes the derivative with respect to $x=\ln{a}$.
Finally, using Eqs. (\ref{phid}), (\ref{gamma}), (\ref{omega}),
(\ref{dec1}) and $\dot{\eta}=1/a$ we obtain
\begin{equation}
\Omega^{'}_{\Lambda}=2\Omega_\Lambda\left\{1+q-\frac{1}{na}\Big(\frac{\Omega_\Lambda}{\gamma_n}\Big)^{1/2}+\Big(\frac{1-\gamma_n}{\gamma_n}+\frac{4\omega\alpha
H^2}{3n^4\phi^2}\frac{\Omega_{\Lambda}}{\gamma_n^2}\Big)\Big[\varepsilon+\frac{1}{na}\Big(\frac{\Omega_\Lambda}{\gamma_n}\Big)^{1/2}\Big]\right\}.\label{motion}
\end{equation}

\subsection{Interacting case}
Our aim here is to construct a cosmological model based on the
Brans-Dicke theory of gravity and on the assumption that the
pressureless DM and ECNADE do not conserve separately but interact
with each other. Since we know neither the nature of DE nor the
nature of DM, a microphysical interaction model is not available
either. However, pressureless DM in interaction with DE is more
reasonable than just another model to describe an accelerated
expansion of the universe. Indeed, this possibility is receiving
growing attention in the literature \cite{Ame} and appears to be
compatible with SNIa and CMB data \cite{Oli}. Interaction causes
the ECNADE and DM do not conserve separately and they must rather
enter the energy balances
\begin{equation}
\dot{\rho}_{\Lambda}+3H(1+w_{\Lambda})\rho_{\Lambda}=-Q,\label{intde}
\end{equation}
\begin{equation}
\dot{\rho}_{m}+3H\rho_{m}=Q,\label{intm}
\end{equation}
\begin{equation}
\dot{\rho}_{r}+4H\rho_{r}=0,\label{eqr2}
\end{equation}
where we have assumed the radiation dose not interact with DE.
Here $Q$ is an interaction term which can be an arbitrary function
of cosmological parameters like the Hubble parameter and energy
densities $Q(H\rho_{\Lambda},H\rho_m)$. The dynamics of
interacting DE models with different $Q$-classes have been studied
in ample detail by \cite{Cabral}. It should be noted that the
ideal interaction term must be motivated from the theory of
quantum gravity. In the absence of such a theory, we rely on pure
dimensional basis for choosing an interaction $Q$. Hence following
\cite{Kim06}, we assume $Q=\Gamma\rho_{\Lambda}$ with
$\Gamma=3b^2(1+u)H$ where $u={\rho_m}/{\rho_{\Lambda}}$ and $b^2$
is a coupling constant. Note that $\Gamma>0$ shows that there is
an energy transfer from the DE to DM. This expression for the
interaction term $Q$ was first introduced in the study of the
suitable coupling between a quintessence scalar field and a
pressureless DM field \cite{Amendola1}.

Using Eqs. (\ref{phid}) and (\ref{critical}), we can rewrite the
first Friedmann equation (\ref{BD1}) as
\begin{equation}\label{rhos}
\rho_{\rm cr}+\rho_k=\rho_r+\rho_m+\rho_{\Lambda}+\rho_{\phi},
\end{equation}
where we have defined
\begin{equation}
\rho_\phi=\frac{1}{2}\varepsilon
H^2\phi^2\Big(\varepsilon-\frac{3}{\omega}\Big).\label{rhophi}
\end{equation}
Dividing Eq. (\ref{rhos}) by $\rho_{\mathrm{cr}}$, this equation
further can be rewritten as
\begin{equation}
\Omega_r+\Omega_m+\Omega_\Lambda+\Omega_\phi=1+\Omega_k,\label{fridmann}
\end{equation}
where
\begin{equation}
\Omega_\phi=\frac{\rho_\phi}{\rho_{\rm
cr}}=-2\varepsilon(1-\frac{\varepsilon\omega}{3}).\label{omegabd}
\end{equation}
Therefore, the interaction term $Q$ can be expressed as
\begin{equation}
Q=3b^2H\rho_\Lambda\Big[\frac{1+\Omega_k-\Omega_r+2\varepsilon(1-\frac{\varepsilon\omega}{3})}{\Omega_\Lambda}\Big].\label{Q}
\end{equation}
Finally, inserting Eqs. (\ref{rhod}) and (\ref{Q}) in Eq.
(\ref{intde}) we find the EoS parameter for the interacting ECNADE
model in the framework of Brans-Dicke theory
\begin{eqnarray}
w_\Lambda=-1-\frac{2\varepsilon}{3}\frac{1}{\gamma_n}+\frac{2}{3na}\Big(2-\frac{1}{\gamma_n}\Big)\Big(\frac{\Omega_\Lambda}{\gamma_n}\Big)^{1/2}
-\frac{8\alpha\omega
H^2}{9n^4\phi^2}\frac{\Omega_\Lambda}{\gamma_n^2}\Big[\varepsilon+\frac{1}{na}
\Big(\frac{\Omega_\Lambda}{\gamma_n}\Big)^{1/2}\Big]\nonumber\\-b^2\Omega_\Lambda^{-1}
\Big[1+\Omega_k-\Omega_r+2\varepsilon\Big(1-\frac{\varepsilon\omega}{3}\Big)\Big].\label{eos2}
\end{eqnarray}
Comparing Eq. (\ref{eos2}) with (\ref{eos1}) shows that in the
presence of interaction since the last expression in Eq.
(\ref{eos2}) has a negative contribution, hence crossing the
phantom divide, i.e. $w_\Lambda<-1$, can be more easily achieved
for than when the interaction between DE and DM is not considered.

In the absence of correction terms ($\alpha=\beta=0$), from Eq.
(\ref{gamma}) we have $\gamma_n=1$ and Eq. (\ref{eos2}) in the
absence of radiation recovers the EoS parameter of interacting
NADE in Brans-Dicke theory \cite{Sheykhi2}
\begin{eqnarray}
w_{\Lambda}=-1-\frac{2\varepsilon}{3}+\frac{2}{3na}\sqrt{\Omega_\Lambda}-b^2
{\Omega^{-1}_\Lambda}\left[1+\Omega_k+2\varepsilon\left(1-\frac{\varepsilon\omega}{3}\right)\right]\label{wDoInt}.
\end{eqnarray}
On the other hand, when $\varepsilon=0$
($\omega\rightarrow\infty$) the Brans-Dicke scalar field becomes
trivial, i.e. $\phi^2=\omega/2\pi G=4\omega M_P^2$, and Eq.
(\ref{eos2}) in the absence of radiation reduces to its respective
expression in ECNADE model in Einstein gravity \cite{karami3}
\begin{eqnarray} w_{\Lambda} =
-1 + \frac{2}{3na}\Big(2 -\frac{1}{\gamma_n}-\frac{\alpha
H^2}{3{M^{2}_P}n^4}\frac{\Omega_{\Lambda}}{\gamma_n^2}\Big)\Big({\frac{\Omega_{\Lambda}}{\gamma_n}}\Big)^{1/2}-
b^2\Big(\frac{1 +
\Omega_k}{\Omega_{\Lambda}}\Big).\label{state-parameter}
\end{eqnarray}
If we compare Eq. (\ref{eos2}) with Eq. (\ref{state-parameter}) we
find out that when ECNADE is combined with Brans-Dicke field the
transition from normal state where $w_\Lambda >-1 $ to the phantom
regime where $w_\Lambda <-1 $ for the EoS of interacting DE can be
more easily achieved for than when resort to the Einstein field
equations is made.

The deceleration parameter $q$ is still obtained according to Eq.
(\ref{dec2}), where $w_{\Lambda}$ is now given by Eq.
(\ref{eos2}). Also the equation of motion for $\Omega_{\Lambda}$
takes the form (\ref{motion}), where $q$ is the deceleration
parameter for the interacting case.

In the end, we would also comment on the resolution of cosmic
coincidence problem, namely, why are the DM and DE densities of
precisely the same order today? In other words, why the ratio of
the two energy densities is of order unity i.e.
$u\equiv\rho_m/\rho_\Lambda\simeq u_0$? (where $u_0$ is a finite
constant of order unity) To show this, we follow the procedure
given in \cite{qui}. Differentiating $u$ w.r.t. $t$ yields
\begin{equation}\label{rdot}
\dot u=3Hu\Big[ w_{\Lambda}+\frac{1+u}{u}\frac{\Gamma}{3H}
\Big]\equiv f(u).
\end{equation}
To find the critical point, we put $f(u)=0$ to get
\begin{equation}\label{cp}
u_c=-\frac{\frac{\Gamma}{3H}}{w_{\Lambda}+\frac{\Gamma}{3H}}.
\end{equation}
Now the cosmic coincidence problem is alleviated if the given
critical point is stable i.e. $f'(u_c)<0$. In other words, a
constraint is obtained on the state parameter
\begin{equation}
w_{\Lambda}<-\frac{\Gamma}{3H}.
\end{equation} Therefore $\Gamma>0$ (refers to transfer of energy from DE to
DM). We emphasize that this result holds independently to any form
of DE and the gravity theory. Moreover the above result holds only
if the background geometry is FRW. However to resolve the
coincidence problem, we need to show that $u_c$ is of order unity.
This will hold if $b^2\sim1/4$ i.e. for a small but positive
coupling parameter, which is also compatible with recent
observational studies \cite{wang1}. We would also comment that this
problem is not resolved by other cosmological models (see e.g.
\cite{campo}).

\section{Conclusions}

Among various candidates to explain cosmic accelerated expansion,
only NADE and HDE models are based on the entropy-area relation.
This implies that the energy density of NADE depends on the length
scales taken in the model. For different choices of length scales,
the dynamics of NADE will be different. Consequently only those
length scales are relevant to DE models which explain certain
physical phenomena like DE dynamical state equation and phantom
crossing. We have chosen the length scale to be the conformal age
of the universe motivated by some previous studies. Note that the
entropy-area relation depends on the gravity theory. When applying
the curvature corrections to the gravity theory, it yields quantum
corrections to the entropy-area relation. Consequently the
definition of NADE acquires additional terms. These correction
terms are important to the DE model when the chosen length scale
is large or small. In the present model, small length scales
yields new physics of the NADE.

We studied an interaction between DE and DM in the Brans-Dicke
framework. The exact nature of this interaction is not understood
due to our ignorance of the microphysics of both components. The
interaction could be exotic and to be explained via beyond the
standard model of particle physics theories. The prime motivation
behind the assumed interaction is to resolve the
cosmic-coincidence problem which asks why
$\Omega_m\sim\Omega_\Lambda$ happens to be at present time? If the
two energy densities match approximately, it suggests emphatically
that the two components evolved not independently. In some recent
studies, the radiation component also has been added to such
interaction models to resolve the cosmic-triple-coincidence
problem \cite{triple} with some new interesting dynamical and
thermodynamical implications are observed.

In an earlier study \cite{Setare2}, the author developed a
correspondence between the HDE and the Brans-Dicke scalar field
and showed that phantom divide is not possible in the Brans-Dicke
gravity. However if the field is assumed to interact with matter
(chameleon scalar field) then the phantom crossing is possible in
the Brans-Dicke framework \cite{jamil1}. Another study
\cite{Sheykhi2} showed that phantom crossing is permissible in the
Brans-Dicke framework if the HDE is replaced with the NADE. In the
present paper, we have extended the later study by incorporating
the correction terms in the NADE definition. An important
consequence/prediction of the present model is that it allows the
phantom crossing of the DE state parameter due to the presence of
several free parameters. Note that the numerical values of
parameters are not entirely arbitrary but are of order unity,
consequently, sub-negative and super-negative values of
$w_\Lambda$ are allowed for certain choices of these parameters.
We emphasize that there is only a hint in some observational
studies on the possible evolution of the equation of state
parameter. Up to know, in spite of its shortcomings on the
theoretical side, the $\Lambda$CDM model appears observationally
more solid than any other in the market \cite{referee}.

\begin{acknowledgments}
The works of K. Karami and A. Sheykhi have been supported
financially by Research Institute for Astronomy $\&$ Astrophysics of
Maragha (RIAAM), Maragha, Iran. M. Jamil would like to thank the
kind hospitality of the Abdus Salam ICTP, Trieste, Italy where part
of this work was completed. We would like to thank E. Abdalla and
the referee for giving enlightening comments to improve this work.
\end{acknowledgments}

\end{document}